\documentclass[12pt,a4paper]{article}
\usepackage{amsmath}
\usepackage{amsfonts} 
\usepackage{graphicx}
\usepackage{geometry}
\usepackage{algorithmicx}
\usepackage{algorithm}
\usepackage{amsthm}

\usepackage{authblk}
\usepackage{hyperref} 
\newtheorem{thm}{Theorem}
\title{Counting Euclidean embeddings of rigid graphs}
\author{Ioannis Z. Emiris}
\author{Ioannis D. Psarros}
\affil{National and Kapodistrian University of Athens, Greece}
\affil{Dept. of Informatics and Telecommunications}

\begin{document}
\maketitle

\begin{abstract}
A graph is called (generically) rigid in $\mathbb{R}^d$ if, for any choice of sufficiently generic edge lengths,
it can be embedded in $\mathbb{R}^d$ in a finite number of distinct ways, modulo rigid transformations.
Here we deal with the problem of determining the maximum number of planar Euclidean embeddings as a function of the number of the vertices.
We obtain polynomial systems which totally capture the structure of a given graph, by exploiting distance geometry theory. Consequently, counting the
 number of Euclidean embeddings of a given rigid graph, reduces to the problem of counting roots of the corresponding polynomial system.

\end{abstract}

\section{Introduction}
Given a graph $G=(V,E)$ with $|V|=n$ and a collection of edge lengths $d_{ij} \in \mathbb{R}^{+}$, a Euclidean embedding of $G$ in $\mathbb{R}^d$ is a mapping of its vertices to a set of points $p_1,\ldots, p_n$, such that $d_{ij} = \Vert p_i - p_j \Vert$, for all $\{i,j\} \in E$. We call a graph (generically) rigid in $\mathbb{R}^d$ iff, for generic edge lengths, it can be embedded in $\mathbb{R}^d$ in a finite number of ways, modulo rigid transformations (translations and rotations). A graph is minimally rigid in $\mathbb{R}^d$ iff it is no longer rigid once any edge is removed. The main goal is to determine the maximum number of distinct planar Euclidean embeddings of minimally rigid graphs, up to rigid transformations, as a function of the number of vertices.

A graph $G=(V,E)$ is called Laman if $|E|=2 |V| -3$ and, additionally, all of its vertex-induced
subgraphs with $3 \leq k < |V|$ vertices to have less than $2k - 3$ edges. It is a fundamental theorem that the class of Laman graphs coincides with the generically minimally rigid graphs in the plane.

A general upper bound on the number of embeddings, for the planar case, is $\binom{2n-4}{n-2}\approx \frac{4^{n-2}}{\sqrt{\pi (n-2)}}$.
The bound was obtained by exploiting results from complex algebraic geometry that bound the degree of (symmetric) determinantal varieties defined by distance matrices \cite{Bor,BorStr,HarTu}.
In \cite{SteThe}, mixed volumes (cf. Section 1.2) also yield an upper bound of $4^{n-2}$, for the planar case.

Here, and in \cite{thesis}, we present an approach, based on distance geometry, which was first proposed in \cite{EmiMor} and \cite{EmTsVa1} for finding an upper bound on the number of Euclidean embeddings modulo rigid transformations.
This approach was used in order to determine an upper bound for the case
of Laman graphs with $7$ vertices~\cite{EmiMor}, which turns out to
be tight.
The same method led to a quite tight bound for the case $n=8$
in~\cite{EmirDesp}.  
However, the approach, as applied in both cases, was not entirely correct.
Our contribution is to rectify both proofs in the following sections.
Interestingly, the upper bound for $n=7$ is correct.  

In Table~\ref{upb1}, we list the known upper bounds
on the number of embeddings of rigid graphs with up to $n=10$ vertices,
also presented in~\cite{EmTsVa1}.
These results constitute an improvement over the results
in~\cite{EmTsVa2}, as we now include the tight bound for 
$n= 7$.  
The bounds for $n=9,10$ are simply multiples by~4 of the respective
bounds for $n-1$.

\begin{table}[H] 
\begin{center} 
\begin{tabular}{|c|c|c|c|c|c|c|c|c|} 
\hline
Number of Vertices &3 & 4& 5& 6 & 7 & 8 & 9 & 10\\
\hline
Upper Bound &2 &4 &8 & 24 & 56 & 128 & 512 & 2048\\
\hline
\end{tabular}
\end{center}
\caption{Upper bounds\label{upb1}}
\end{table}

\section{Laman graphs and Henneberg steps}

We will use a combinatorial characterization of rigidity in $\mathbb{R}^2$, which is based on Laman graphs \cite{Lam,Max}.
A graph $G=(V,E)$ with $|V|=n$ is called Laman iff $|E|=2n-3$ and all of its induced subgraphs on $k<n$ vertices have $\leq 2k-3$ edges.
It is a fundamental theorem (Maxwell 1864 - Laman 1970) that a graph is minimally rigid in $\mathbb{R}^2$ iff it is a Laman graph.
Moreover, we know that Laman graphs admit inductive constructions that begin with a triangle, followed by a sequence of so-called Henneberg steps (Henneberg constructions). A graph is Laman iff it has a Henneberg construction.
Each Henneberg step adds to the graph a new vertex and a total number of two edges.

A Henneberg-$1$ (or $H_1$) step connects the new vertex to two existing vertices and it doubles the number of Euclidean embeddings.

\begin{figure}[H]
  \centering
                \includegraphics[scale=0.7]{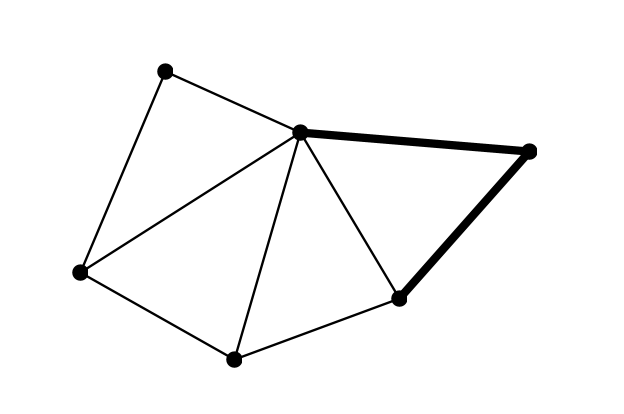}
  \caption[]
   {$H_1$ step}
\end{figure}

A Hennenerg-$2$ (or $H_2$) step connects the new vertex to three existing vertices having at least one edge among them, and this edge is removed.

\begin{figure}[H]
  \centering
                \includegraphics[scale=0.7]{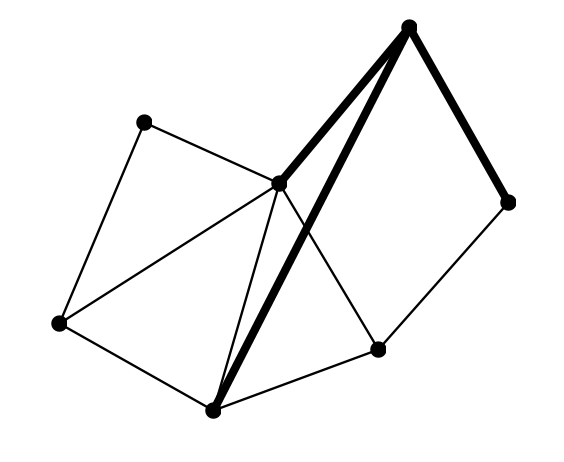}

  \caption[]
   {$H_2$ step}
\end{figure}
A Laman graph is called $H_1$ if it can be constructed using only $H_1$ steps and it is called $H_2$ otherwise.

\section{Polynomial systems and mixed volume}
Given a polynomial $f$ in $n$ variables, its support is the set of exponents in $N^n$ corresponding to nonzero terms. The Newton polytope of $f$ is the convex hull of its support and lies in $R^n$. Consider the Minkowski sum of the scaled polytopes
$\lambda_1 P_1 + · · · + \lambda_n P_n \in R^n $. Minkowski's theorem states that its n-dimensional Euclidean volume is a polynomial in $\lambda_1,..,\lambda_n$ homogeneous, of degree n. The mixed volume is the coefficient of the monomial $\lambda_1 \cdot \cdot \cdot \lambda_n$.
 \begin{thm}
  Let $f_1 =\cdots = f_n = 0$ be a polynomial system in n variables with real coefficients, where the $f_i$ have
fixed supports. Then, the number of common isolated solutions in $(C∗ )^n$ is bounded above by the mixed volume of (the
Newton polytopes of ) the $f_i$ . This bound is tight for a generic choice of coefficients of the $f_i$ ’s.

 \end{thm}

\section{Distance Geometry}
Given a graph with $n$ vertices, the corresponding pre-distance matrix is a symmetric $n\times n$ matrix $D$, where $D_{ij}=d_{ij}^2$.
The corresponding Cayley-Menger matrix $CM(D)$ is a symmetric $(n+1)\times (n+1)$ matrix $B$ where $B_{0i}=1$ , $B_{i0}=1$ for $0<i\leq n$ and $B_{ij}=D_{ij}$ for $1\leq i,j\leq n$.

\begin{thm}
Let $B$ be a Cayley-Menger matrix as defined before. Then
$B$ corresponds to a graph embeddable in $\mathbb{R}^{d}$ if
\begin{itemize}
\item rank$(B) \leq d+2$,
\item $(-1)^k D(i_1,i_2,\dots,i_k)\geq 0$, $\forall k \geq 2$,
\end{itemize}
where $D(\cdot)$ is any minor of $B$ indexed by $0,i_1,i_2,\dots,i_k$.
\end{thm}

In other words, $D(\cdot)$ must also be Cayley-Menger.
The second condition yields inequalities. For instance, for $k=2$ it yields the fact that all distances must be positive and for $k=3$ it yields the triangular inequality. The first condition implies the vanishing of all $(d+3) \times (d+3)$ Cayley-Menger minors. We use this condition in order to construct well-constrained square systems whose roots correspond to Euclidean Distance Matrix Completions of $B$.
Note that these systems do not contain all of the unknown variables but they uniquely define the ones not contained.

 Hence, an upper bound in the number of roots is equivalent to an upper bound in the number of Euclidean embeddings modulo rigid transformations and reflection. Multiplying by $2$ yields an upper bound in the number of Euclidean embeddings modulo rigid transformations. This fact was not taken into consideration in \cite{EmiMor,EmTsVa1}. However, the upper bounds obtained here are actually the same with \cite{EmiMor,EmTsVa1}, a fact that lead us to the conjecture that there exist systems that count the exact number of Euclidean embeddings modulo rigid transformations.  

\section*{The case $n=6$}

We apply the aforementioned
approach in the Desargues framework which is known to have $24$ Euclidean embeddings in the plane.

\begin{figure}[H] \centering
  \includegraphics[width=0.5 \textwidth]{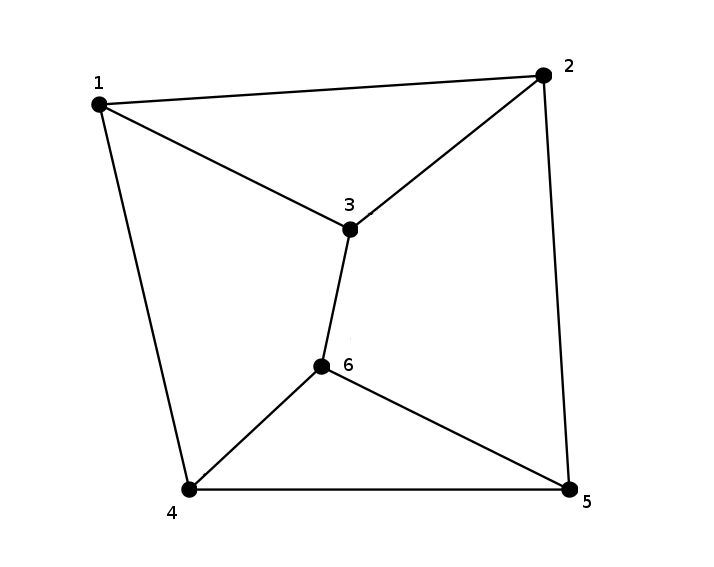}
\caption{Desargues Graph}
\end{figure}

The corresponding Cayley-Menger Matrix is the following:
$$
\begin{bmatrix}
0 & 1 & 1 & 1 & 1 & 1 & 1 \\
  1 & 0 & c_{12} & c_{13} & c_{14} & x_{15} & x_{16}    \\ 
  1 & c_{12} & 0 & c_{23} & x_{24} & c_{25} & x_{26}  \\
  1 & c_{13} & c_{23} & 0 & x_{34} & x_{35} & c_{36} \\
  1 & c_{14} & x_{24} & x_{34} & 0 & c_{45} & c_{46}  \\
  1 & x_{15} & c_{25} & x_{35} & c_{45} & 0 & c_{56}  \\
  1 & x_{16} & x_{26} & c_{36} & c_{46} & c_{56} & 0

 \end{bmatrix}
$$

We are now able to find at least one subsystem that satisfies the laman property and hence it has finitely many roots. The following $3 \times 3$ system also uniquely defines the rest of the unknown variables.
$$
D(1, 4, 5, 6)(c_{14}, c_{45}, c_{46}, c_{56}, x_{15}, x_{16})=0$$$$
D(1, 3, 5, 6)(c_{13}, c_{36}, c_{56}, x_{15}, x_{16}, x_{35})=0$$$$
D(1, 2, 3, 5)(c_{12}, c_{13}, c_{23}, c_{25}, x_{15}, x_{35})=0
$$
This system has mixed volume equal to $12$. In other words, it yields at most $12$ complex roots. Hence, $12$ is an upper bound on the Euclidean embeddings modulo rigid transformations and reflection. Taking symmetry into account, we have at most 24 Euclidean embeddings modulo rigid transformations.

Another case of interest is that of the complete bipartite graph with $6$ vertices.
\begin{figure}[H] 
 \centering
  \includegraphics[width=0.5 \textwidth]{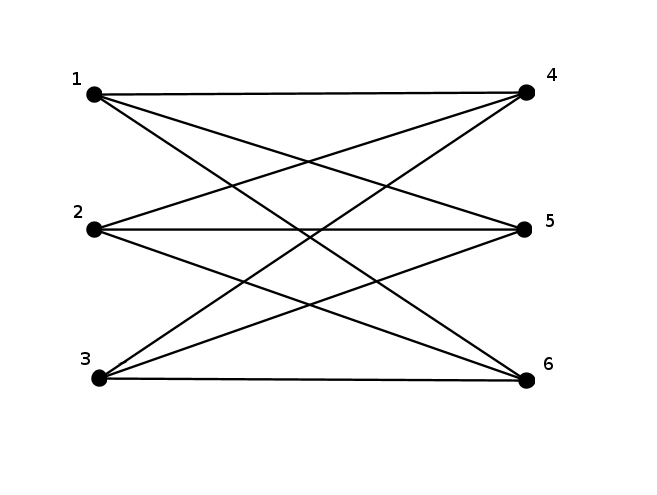}
\caption[]
   {$K_{3,3}$}
\end{figure}
The corresponding Cayley-Menger matrix is the following:
$$
\begin{bmatrix}
0 & 1 & 1 & 1 & 1 & 1 & 1 \\
  1 & 0 & x_{12} & x_{13} & c_{14} & c_{15} & c_{16}    \\ 
  1 & x_{12} & 0 & x_{23} & c_{24} & c_{25} & c_{26}  \\
  1 & x_{13} & x_{23} & 0 & c_{34} & c_{35} & c_{36} \\
  1 & c_{14} & c_{24} & c_{34} & 0 & x_{45} & x_{46}  \\
  1 & c_{15} & c_{25} & c_{35} & x_{45} & 0 & x_{56}  \\
  1 & c_{16} & c_{26} & c_{36} & x_{46} & x_{56} & 0

 \end{bmatrix}
$$ 
We choose the following system which satisfies the abovementioned properties:
$$
D(1, 2, 3, 6)(c_{16}, c_{26}, c_{36}, x_{12}, x_{13}, x_{23})=0$$$$
D(1, 2, 3, 5)(c_{15}, c_{25}, c_{35}, x_{12}, x_{13}, x_{23})=0$$$$
D(1, 2, 3, 4)(c_{14}, c_{24}, c_{34}, x_{12}, x_{13}, x_{23})=0
$$
This system has mixed volume equal to $11$ which yields an upper bound of $22$ in the number of Euclidean embeddings modulo rigid transformations.

\section{The case $n=7$}

This case was first studied in \cite{EmiMor}. However, the property that an embedding and its mirror image preserve distances between points, was not taken into account. However, although we provide a new and correct proof, the same
results are eventually obtained.

Among all Laman graphs, there are only $3$ $H_2$ graphs, we have to consider. All $H_1$ graphs have $2^5=32$ Euclidean embeddings. We can obtain an upper bound of $52$ and $48$ Euclidean embeddings for the two of these graphs. The $H_2$ graph with the highest number of Euclidean embeddings is the following: 
\begin{figure}[H] 
 \centering
  \includegraphics[width=0.5 \textwidth]{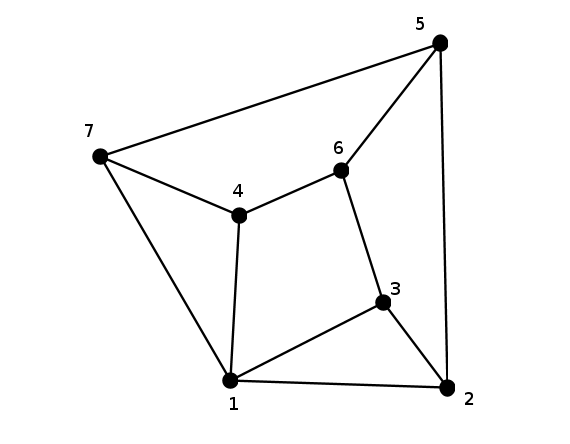}
\caption{Worst case for $n=7$}
\end{figure}
The corresponding Cayley-Menger matrix is the following:
$$
\begin{bmatrix}
0 & 1 & 1 & 1 & 1 & 1 & 1 & 1 \\
  1 & 0 & c_{12} & c_{13} & c_{14} & x_{15} & x_{16} & c_{17}    \\ 
  1 & c_{12} & 0 & c_{23} & x_{24} & c_{25} & x_{26} & x_{27} \\
  1 & c_{13} & c_{23} & 0 & x_{34} & x_{35} & c_{36} & x_{37}\\
  1 & x_{14} & x_{24} & x_{34} & 0 & x_{45} & c_{46}  & c_{47}\\
  1 & x_{15} & c_{25} & x_{35} & x_{45} & 0 & c_{56}  & c_{57}\\
  1 & x_{16} & x_{26} & c_{36} & c_{46} & c_{56} & 0 & x_{67}\\
 1 & c_{16} & x_{26} & x_{36} & c_{46} & c_{56} & x_{67} & 0\\

 \end{bmatrix}
$$ 
We can obtain an upper bound for this graph from the following system.
$$
D(1, 4, 5, 7)(c_{14}, c_{17}, c_{47}, c_{57}, x_{15}, x_{45})=0$$$$
D(1, 3, 5, 6)(c_{13}, c_{36}, c_{56}, x_{15}, x_{16}, x_{35})=0$$$$
D(1, 4, 5, 6)(c_{14}, c_{46}, c_{56}, x_{15}, x_{16}, x_{45})=0$$$$
D(1, 2, 3, 5)(c_{12}, c_{13}, c_{23}, c_{25}, x_{15}, x_{35})=0
$$
Its mixed volume equals to $28$. Hence, we obtain an upper bound of $56$ Euclidean embeddings.

\section{The case $n=8$}

The case of graphs with $8$ vertices was studied in \cite{EmirDesp}.
The authors conclude that the upper bound follows from the upper bound
of $27$ specific $H_2$ graphs.
Unfortunately, the systems used in order to determine the number of embeddings for these $27$ graphs do not satisfy the Laman property. Hence, the final result in \cite{EmirDesp} is not correct. In the following paragraphs, we present the difficulties we faced while trying to correct this error, and we make a conjecture about the actual upper bound on the number of Euclidean embeddings modulo rigid transformations in the case of Laman graphs with 8 vertices.

Notice that the systems which provide us with upper bounds
in the case of graphs with $6$ and $7$ vertices, are of size $(n-3)\times (n-3)$,
where $n$ denotes the number of vertices.
Unfortunately, in the case of graphs with $8$ vertices, there are graphs for which, we can only obtain systems of size $(n-2) \times (n-2)$. The mixed volume of these systems, multiplied by $2$ is not representative of the maximum number of Euclidean embeddings.

 However, experiments show that these systems take into account reflection and count the number of Euclidean embeddings modulo rigid transformations. In other words, we conjecture that we do not have to multiply by $2$, the mixed volume we obtain from some bigger systems. We expect this to happen for general $n$. However, this is only a conjecture.

\begin{figure}[H] 

 \centering

  \includegraphics[width=1 \textwidth]{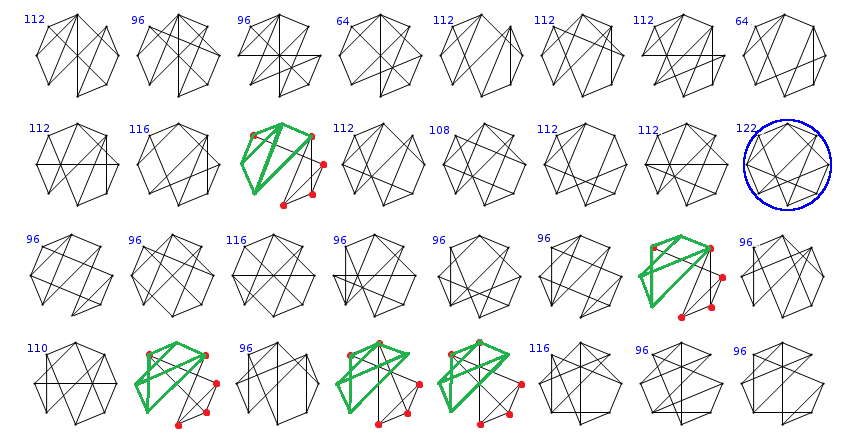}

\caption {Graphs first studied in \cite{EmirDesp}}\label{res8}
\end{figure}

In table ({\ref{res8}}), we present the mixed volumes obtained by systems that are of size $(n-2) \times (n-2)$ and also satisfy the Laman property for the abovementioned $27$ graphs.

Hence, we expect that the upper bound on the number of embeddings for Laman graphs with $8$ vertices is $122$, while the current upper bound is $128$.
\section{Further Work}

The most relative open problem to our work is to understand whether there exist systems that take also into account reflection. If we prove that systems like the ones used for the results of table [\ref{res8}], count also embeddings produced by reflection of other embeddings, we will have an upper bound for the graphs with $8$ vertices.

Undoubtedly, the most important and oldest problem in rigidity theory is the full combinatorial characterization
of rigid graphs in $\mathbb{R}^3$. In the planar case, existing bounds are not tight. This is due to the fact that root counts
include rotated copies of certain embeddings. A related problem is to derive an algorithmic process for obtaining
good algebraic representations, e.g. for producing low mixed volumes. Distance matrices may offer such a general
approach, including the spatial case. For high $n$ the issue is that the number of equations produced is quite large,
with algebraic dependancies among them.

Since we deal with Henneberg constructions, it is important to determine the effect of each step on the number
of embeddings: a $H_1$ step always doubles their number; we conjecture that $H_2$ multiplies it by $\leq 4$ and spatial $H_3$ by
$\leq 8$, but these may not always be tight. Our conjecture has been verified for small $n$.

\end{document}